\documentclass{PoS}

\title{{\footnotesize{DESY 18-160, DO-TH 18/21}}\\
Computer algebra tools for Feynman integrals and related multi-sums\thanks{This
work
was supported in part by the Austrian Science Fund (FWF) grant SFB F50
(F5009-N15)}} 

\ShortTitle{ Computer algebra tools for Feynman integrals and related multi-sums}

\author{Johannes Bl\"umlein\\
Deutsches Elektronen--Synchrotron, DESY, \\
                   Platanenallee 6, D-15738 Zeuthen, Germany\\
Email: \email{Johannes.Bluemlein@desy.de}}
\author{\speaker{Carsten Schnneider}\\
        Research Institute for Symbolic Computation (RISC)\\
        Johannes Kepler University, Altenbergerstra\ss{}e 69, A-4040 Linz,
Austria\\
        E-mail: \email{Carsten.Schneider@risc.jku.at}\\
}

\abstract{In perturbative calculations, e.g., in the setting of Quantum Chromodynamics (QCD) one aims at the 
evaluation of Feynman 
integrals. Here one is often faced with the problem to simplify multiple nested integrals or sums to expressions in 
terms 
of indefinite nested integrals or sums. Furthermore, one seeks for solutions of coupled systems of linear differential equations, that can be represented in terms of indefinite nested sums (or integrals). In this article we elaborate the main tools and the corresponding packages, that we have developed and intensively used within the last 10 years in the course of our QCD-calculations.}

\FullConference{Loops and Legs in Quantum Field Theory (LL2018)\\
		29 April 2018 - 04 May 2018\\
		St. Goar, Germany}

\usepackage{graphicx}

\usepackage{amsmath,amsthm}
\usepackage{amssymb}
\usepackage{epsfig}
\usepackage{etex, xy}
\xyoption{all}

\theoremstyle{definition}
\newtheorem{remark}{Remark}
\newtheorem{example}{Example}
\newtheorem{definition}{Definition}

\usepackage[square,numbers,sort&compress]{natbib}

		\newcommand{\ep}{\varepsilon}

\newcommand{\NN}{\mathbb{N}}
\newcommand{\ZZ}{\mathbb{Z}}

\newcommand{\KK}{\mathbb{K}}

\newcommand{\set}[1]{{\mathbb #1}}

\newcommand{\ProblemSpec}[1]{\noindent\fcolorbox[rgb]{0,0,0}{0.7,0.9,0.7}{\begin{minipage}{14.8cm}
                                 #1
                                \end{minipage}}}

\def\href#1#2{#2}
		
\begin{document}

\section{Introduction}

We aim at the simplification of loop Feynman integrals to expressions in terms of special functions and constants. 
More precisely, we consider $\mu$-loop massive Feynman parameter integrals (e.g., $\mu=3$) 
emerging in renormalizable Quantum Field Theories, like
Quantum Electrodynamics or Quantum Chromodynamics, see e.g. \cite{BKSF:12,WeinzierlBook:13}, which can be transformed to a linear combination of $s$-fold multiple integrals of the form
\begin{equation}\label{Equ:GeneralForm}
F(m_1,\dots,m_r,n,\ep)=\int_0^1\dots\int_0^1f(m_1,\dots,m_r,n,\ep,x_1,\dots,x_s)dx_1\dots\,dx_s;
\end{equation}
$m_1,\dots,m_r$ are the involved masses, the discrete parameter $n$ stands for the Mellin variable, and $D:=4+\ep$ with $\ep\in\set R$ is the analytic continuation of the space-time dimension. 
Often one can consider $m_1,\dots,m_r$ and $\ep$ as indeterminates (and reinterprets them, only if necessary, as concrete parameters that can be evaluated within certain ranges). In the following we assume that the masses are contained in a field $\KK$ of characteristic $0$. A crucial property is that the integrand $f$ of such a Feynman integral is hyperexponential\footnote{$h(x)$ is hyperexponential (resp.\ hypergeometric) in $x$ if $\frac{h'(x)}{h(x)}$ (resp.\ $\frac{h(x+1)}{h(x)}$) is a rational function in $x$.} in each of the integration variables $x_i$ $(1\leq i\leq s)$ and hypergeometric in the discrete parameter $n$.

Given such an integral\footnote{In the following we will suppress the mass dependencies.} $F(n,\ep)$, one seeks for the first coefficients of their Laurent series expansion 
$$F(n,\ep)=F_l(n)\ep^l+F_{l+1}(n)\ep^{l+1}+\dots+F_r(n)\ep^r+O(\ep^{r+1})$$
where the coefficients are given in terms of special functions and constants; usually, we have $l=-\mu$ for a $\mu$-loop Feynman integral and $r=0,1,2$.
More precisely, we applied the following general strategies in the course of our calculations:
$$\xymatrix@!R=1cm@C2.cm{
\ar[dd]_(.5){\txt{\footnotesize symbolic integration}}\txt{Feynman integral}\ar@/^1.5pc/[dr]_(.6){\txt{\footnotesize 
${}_pF_q$-/Mellin-Barnes\\\footnotesize technologies}}&\\
&\txt{multi-sum expressions}\ar@/^2pc/[dl]_(.35){\txt{\footnotesize symbolic summation}}\\
\txt{special function expressions}&
}$$
First one may apply directly symbolic integration algorithms (see Section~\ref{Sec:Integration}) in combination with recurrence solving technologies (see Section~\ref{Sec:RecurrenceSolving}) to simplify these integrals in terms of special functions. Second one can transform in a preprocessing step these integrals to expressions in terms of multiple nested sums. Namely, by applying successively Newton's binomial theorem and Mellin-Barnes decompositions on the integrand, implemented in different packages 
\cite{Czakon:2005rk,Smirnov:2009up,Gluza:2007rt}, one can carry out all integrals by introducing Mellin-Barnes integrals. Finally, applying the residue theorem to these introduced Mellin-Barnes integrals yields multiple sums over hypergeometric expressions. We emphasize that this mechanism has to be applied very carefully in order to arrive at expressions that are suitable for our symbolic summation methods.
In general, we will end up at a linear combination of hypergeometric multi-sums of the form
\begin{equation}\label{Equ:SumRep}
S(n,\ep)=
\sum_{k_1=1}^{L_1(n)}\dots\sum_{k_v=1}^{L_v(n,k_1,\dots, k_{v-1})} 
f(n,\ep,k_1,\dots,k_v)
\end{equation}
over $\KK(n,\ep)$.
Here the upper bounds $L_1(n),\dots,L_{v}(n,k_1,\dots,k_{v-1})$ are integer linear (i.e., linear combinations of the variables over the integers) in
the dependent parameters or $\infty$, and $f$ is hypergeometric in $n$ and the summation variables $k_i$. 
Then given such a sum representation, we are in the position to apply our symbolic summation tools summarized in Section~\ref{Sec:Summation}.

In many applications, one is faced with thousands (even millions) of Feynman integrals that describe an underlying physical problem. To treat them directly with the above methods is usually out of scope. As a preprocessing step, one utilizes integration by parts (IBP) methods~\cite{Chetyrkin:1981qh,Laporta:2001dd,Studerus:2009ye,vonManteuffel:2012np,MARSEID}. They crunch the given expression to a more compact form in terms of only few integrals, that have to be treated individually. Similarly, while transforming Feynman integrals to multiple sums, one obtains enormous expressions consisting of 
up to thousands of multiple sums. To simplify these sums successively using these summation tools is not 
only problematic because of time limitations, but also because of the following intrinsic problem. 
Often the 
scattered sums themselves cannot be simplified in terms of indefinite nested sums, but only a suitable combination of them can be simplified in this way. In order to bypass these problems, the package \texttt{SumProduction}~\cite{Schneider:12a,Schneider:13b} built on \texttt{Sigma}~\cite{Schneider:07a,Schneider:13a} can be utilized. It reduces the sum expressions to compact forms where the arising sums are merged appropriately. Afterwards the symbolic summation can be applied to these expressions within reasonable time and without dealing with sums that cannot be handled within the difference 
ring setting~\cite{DR1,DR2,DR3}. Summarizing, the following simplification techniques are applied in addition in order to reduce the given problem to a reasonable size of integrals or sums. 
$$\xymatrix@!R=1.cm@C0.2cm{
&\ar@/^-1.5pc/[dl]_(0.6){\txt{\footnotesize IBP methods}}\txt{huge expression of\\ Feynman integrals}\ar@/^1.5pc/[dr]^(0.75){\txt{\footnotesize ${}_pF_q$-/Mellin-Barnes\\\footnotesize technologies}}&\\
\txt{huge expression in terms of\\ few master integrals}&&\txt{huge expression\\ of (small) multi-sums}\ar[d]_{\txt{\footnotesize merging/reduction}}\\
&&\txt{huge expression in terms\\ of few (large) multi-sums}
}$$
An extra advantage of the IBP approach is that most of the produced master integrals are described as solutions of coupled systems of linear differential equations. Only few integrals (usually of simpler type) arise in the inhomogeneous part of these systems which themselves are not represented as solutions of coupled systems. 
Thus the remaining task is to simplify these few integrals by symbolic summation/integration methods and to derive the remaining (and usually more complicated) integrals by solving the provided coupled systems of linear differential equations.

In a nutshell, one is faced with the problem to simplify Feynman integrals by means of symbolic integration (see Section~\ref{Sec:Integration} in combination with Section~\ref{Sec:RecurrenceSolving}), symbolic summation (see Section~\ref{Sec:Summation}) or solving coupled systems of linear differential equations (see Section~\ref{Sec:CoupledSystem}).


\section{Computer algebra methods}

In the following we will present the most relevant computer algebra tools, that were crucial to carry out the challenging QCD-calculations described in~\cite{CoupledSys1,CoupledSys2,CoupledSys3,Schneider:16b,CoupledSys4,CoupledSys5,Physics1,Physics2}. For a comprehensive summary of further available tools we refer to the recent survey article~\cite{Blumlein:2018cms}.

\subsection{Symbolic integration: the multivariate Almkvist-Zeilberger algorithm}\label{Sec:Integration}

Concerning symbolic integration tools we heavily utilized the multivariate Almkvist-Zeil\-berger algorithm~\cite{AZ:90,AZ:06}, in particular an optimized and improved version~\cite{Ablinger:12,Schneider:16b} for Feynman integrals that can tackle the following problem.

\medskip

\ProblemSpec{\noindent \textbf{Recurrence finding.}\\
\textit{Given} an integral in the form~\eqref{Equ:GeneralForm} 
where $f$ is hyperexponential in $x_i$ for $i=1,2,\dots,s$ and hypergeometric in $n$.
\textit{Compute} a linear recurrence of the form
\begin{equation}\label{Equ:RecEp}
 a_0(n,\ep)F(n,\ep)+a_1(n,\ep)F(n+1,\ep)+\dots+a_d(n,\ep)F(n+d,\ep)=0
\end{equation}
with polynomials $a_i(n,\ep)\in\KK[\ep,n]$ where $a_d(n,\ep)\neq0$.}

\medskip

\noindent Internally, the problem is reduced to solve linear systems of equations in $\KK(n,\ep)$. The complexity to solve the underlying systems depends heavily on the involvement of the constructed field $\KK$, the number of the integration variables $x_1,\dots,x_s$ and on the structure of the integrand $f$ (in particular how the variables are intertwined). However, for various concrete situations this method works very well.

\begin{example}\label{Exp:MAZ}
 If one applies the package \texttt{MultiIntegrate}~\cite{Ablinger:12,Schneider:16b} to the multi-integral
\begin{equation}\label{Equ:QuadrupleIntegral}
F(n,\ep)=\int_0^1\int_0^1\int_0^1\int_0^{1-u}\tfrac{
(x+y-1)^n
x^{\varepsilon/2} (1-x)^{\varepsilon/2} 
y^{\varepsilon/2} (1-y)^{\varepsilon/2} 
(1-u-v)^{n}\big(1-u \tfrac{x}{x-1}-v \tfrac{y}{y-1}\big)^{-1+3/2\varepsilon} }{u^{1+\varepsilon/2} v^{1+\varepsilon/2}}
dx\,dy\,du\,dv,
\end{equation}
one can compute a recurrence of the form~\eqref{Equ:RecEp} with order $d=5$ in about 8 hours.
\end{example}

\begin{remark}
 For some special cases we could also utilize an extended version~\cite{vonManteuffel:2014qoa} of the hyperlogarithm method \cite{Brown:2008um,Panzer:2014caa} in order to tackle massive Feynman integrals. 
\end{remark}

\subsection{Recurrence solving}\label{Sec:RecurrenceSolving}

Suppose we calculated a recurrence of the form~\eqref{Equ:RecEp} or more generally of the form
\begin{equation}\label{Equ:RecNoEp}
 a_0(n)F(n)+a_1(n)F(n+1)+\dots+a_d(n)F(n+d)=b(n)
\end{equation}
with $a_i(n)\in\KK[n]$ for $0\leq i\leq d$ where $a_d(n)\neq0$ and 
where $b(n)\in\KK$ for $n\geq0$. Let $\delta=\max(\{i\in\NN\mid a_d(i)=0\}\cup\{-1\})+1$. Then it follows immediately that there is a unique sequence solution $F(n)$ with $n\geq\delta$ of~\eqref{Equ:RecNoEp} if one fixes the first $d$ initial values $F(\delta),\dots,F(\delta+d-1)\in\KK$.\\
An important task is then to represent, if possible, such a solution in terms of special functions. Within the summation package~\texttt{Sigma}~\cite{Schneider:07a,Schneider:13a} difference ring algorithms have been encoded~\cite{Karr:81,Abramov:94,Singer:99,Schneider:01,Schneider:05c,Schneider:07d,Schneider:08c,Petkov:10,Schneider:10b,Schneider:15,DR1,DR2,DR3},
that find such a representation in terms of indefinite nested sums over hypergeometric products, whenever this is possible.

\begin{definition}
Let $\KK$ be a field of characteristic $0$. A product $\prod_{j=l}^kf(j)$, $l\in\NN$, is called \textit{hypergeometric} in $k$ over $\KK$ if $f(y)$ is an element from the rational function field $\KK(y)$ where the numerator and denominator of
$f(j)$ are nonzero for all $j\in\ZZ$ with $j\geq l$.
An expression in terms of \textit{indefinite nested sums over hypergeometric products in $k$ over $\KK$} is composed recursively by the three operations 
($+,-,\cdot$) with
\begin{itemize}
\item elements from the rational function field $\KK(k)$,
\item hypergeometric products in $k$ over $\KK$,
\item and sums of the form $\sum_{j=l}^kf(j)$ with $l\in\NN$ where 
$f(j)$ is an expression in terms of indefinite nested sums over hypergeometric products in $j$ over $\KK$; here it is assumed that the evaluation of $f(j)|_{j\mapsto\lambda}$ for all $\lambda\in\ZZ$ with $\lambda\geq l$ does not introduce any poles.
\end{itemize}
If $\KK$ and $k$ are clear from the context we simply say that $f(k)$ is an expression in terms of \textit{indefinite nested sums (over hypergeometric products)}.
\end{definition}

\begin{example}
 The class of indefinite nested sums over hypergeometric products in $n$ over $\KK$ covers as special cases harmonic sums~\cite{Bluemlein:99,Vermaseren:99} like
$$S_{2,1}(n)=\sum_{i=1}^n\frac{1}{i^2}\sum_{j=1}^i\frac{1}{j},$$
generalized harmonic sums~\cite{Moch:02,ABS:13} like
$$\sum_{k=1}^n\frac{2^k}{k}\sum_{i=1}^k \frac{\displaystyle2^{-i}}{i}\sum_{j=1}^i \frac{1}{j},$$
cyclotomic harmonic sums~\cite{ABS:11} like
$$\sum_{k=1}^n \frac{1}{(1+2 k)^2}\sum_{j=1}^k \frac{1}{j^2}\sum_{i=1}^j \frac{1}{1+2 i}$$
or nested binomial sums~\cite{ABRS:14} like 
$$\sum_{k=1}^n \frac{1}{\binom{2k}{k}^2}\sum_{j=1}^k \frac{1}{j^2}\sum_{i=1}^j \frac{\binom{2i}{i}}{1+2 i}.$$
\end{example}

\noindent We can solve the following problem with the summation package \texttt{Sigma} for the class of indefinite nested sums.

\medskip

\ProblemSpec{\textbf{Recurrence solving.}\\
\textit{Given} $a_i(n)\in\KK[n]$ for $0\leq i\leq d$ where $a_d(n)\neq0$ and given 
$b(n)$ that can be calculated by an expression in terms of indefinite nested sums over hypergeometric products;
given $\delta$ as above and $c_{\delta},\dots,c_{\delta+d-1}\in\KK$. \textit{Decide} constructively if the solution $F(n)$ of~\eqref{Equ:RecNoEp} with $F(i)=c_i$ for $\delta\leq i\leq \delta+d-1$ can be calculated by an expression in terms of indefinite nested sums over hypergeometric products.
}

\medskip

In QCD-calculations the recurrence~\eqref{Equ:RecNoEp} usually depends on $\ep$. In this case we could treat $\ep$ just as an extra parameter which is contained in $\KK$ and seek for a solution within the class of indefinite nested sums over hypergeometric products over $\KK$. However, in most cases such an all $n$ and $\ep$ representation  does not exist. A more flexible strategy is to hunt for a solution $F(n,\ep)$ given in its $\ep$-expansion
\begin{equation}\label{Equ:LaurentSeriesSol}
F(n,\ep)=F_l(n)\ep^l+F_{l+1}(n)\ep^{l+1}+\dots
\end{equation}
More precisely, consider a recurrence of the form
\begin{equation}\label{Equ:RecEpFull}
 a_0(n,\ep)F(n,\ep)+a_1(n,\ep)F(n+1,\ep)+\dots+a_d(n,\ep)F(n+d,\ep)=b(n,\ep)
\end{equation}
with $a_i(n,\ep)\in\KK[n,\ep]$ where not all\footnote{If all $a_i(n,0)$ are zero, we can divide the equation by $\ep^r$ for some $r\in\NN$ yielding again polynomial coefficients on the left-hand side where not all are zero when setting $\ep$ to $0$.} $a_i(n,0)$ are zero and with a right-hand side given in its $\ep$-expansion
\begin{equation}\label{Equ:RHSExpansion}
b(n,\ep)=\sum_{i=l}^{\infty}b_i(n)\ep^i
\end{equation}
where $b_i(n)\in\KK$ for $i\geq l$ and $n\geq0$.
Let $o$ be the largest integer such that $a_o(n,0)\neq0$ and 
$\delta=\max(\{i\in\NN\mid a_o(i,0)=0\}\cup\{-1\})+1$. 
Then by a slight variation of~\cite{BKSF:12} it follows that there is a unique Laurent series solution~\eqref{Equ:LaurentSeriesSol}
for $n\geq\delta$ when fixing  $F_i(j)\in\KK$ for $i\geq l$ and $\delta\leq i\leq\delta+o$. 
In particular, the following holds. If there are two solutions, say $F(n,\ep)$ with~\eqref{Equ:LaurentSeriesSol} and $F'(n,\ep)=F'_l(n)\ep^l+F'_{l+1}(n)\ep^{l+1}+\dots$ and if $F_i(j)=F'_i(j)$ for $l\leq i\leq r$ and $\delta\leq i\leq\delta+o$, then $F_i(n)=F'_i(n)$ for all $l\leq i\leq r$ and $n\geq\delta$. In short, knowing the first initial values determines uniquely the first coefficients of a Laurent series solution. In particular, they can be prolonged stepwise to a full Laurent series solution by fixing further initial values of higher $\ep$-orders.

\medskip

\noindent With this preparation step we can now introduce the following machinery from~\cite{BKSF:12} implemented in~\texttt{Sigma}.

\medskip

\ProblemSpec{\textbf{Recurrence solving for $\ep$-expansions.}\\
\textit{Given} $a_i(n,\ep)\in\KK[n,\ep]$ for $0\leq i\leq d$ where not all $a_i(n,0)$ are zero and given $b(n,\ep)$ with~\eqref{Equ:RHSExpansion} where 
$b_l(n),\dots,b_r(n)$ are represented in terms of indefinite nested sums over hypergeometric products; given $o,\delta$ as given above, and given $c_{i,j}\in\KK$ for $l\leq i\leq r$ and $\delta\leq i\leq\delta+o$.
\textit{Decide} constructively if the first coefficients $F_l(n),\dots,F_r(n)$ of a Laurent series solution~\eqref{Equ:LaurentSeriesSol} of the recurrence~\eqref{Equ:RecEpFull} with $F_i(j)=c_{i,j}$ can be calculated by expressions in terms indefinite nested sums over hypergeometric products.
}

\begin{example}
We continue Example~\ref{Exp:MAZ} by taking the recurrence of order $5$ (see~\cite[pp.~49--52]{Schneider:16b}) for the integral $F(n,\ep)$ given in~\eqref{Equ:QuadrupleIntegral}. In addition, we take the five initial values $F(2,\ep),\dots,F(6,\ep)$ expanded up to $\ep^{-1}$. E.g., we have
$$F(2,\ep)=\frac{20}{27 \varepsilon^3}
-\frac{40}{27 \varepsilon^2}
+\frac{1}{\ep} \left(\frac{1393}{486}
+\frac{5 \zeta_2}{18}\right)+O(\ep^0).$$
Then using the above algorithm we can calculate the first three coefficients in
$$F(n,\ep)=F_{-3}(n)\,\ep^{-3}+F_{-2}(n)\,\ep^{-2}+F_{-1}(n)\,\ep^{-1}+O(\ep^0)$$
of the $\ep$-expansion within a few seconds. More precisely, we get
\begin{align*}
F_{-3}(n)&=\frac{8 (-1)^n}{3 (n+1) (n+2)}
+\frac{8 (2 n+3)}{3 (n+1)^2 (n+2)},\\
F_{-2}(n)&=-\frac{4(-1)^n\big(
        3 n^3+18 n^2+31 n+18\big)}{3(n+1)^3 (n+2)^2} 
-\frac{4\big(
        6 n^3+32 n^2+51 n+26\big)}{3(n+1)^3 (n+2)^2},\\
F_{-1}(n)&=(-1)^n \left(
        \frac{2\big(
                9 n^5+81 n^4+295 n^3+533 n^2+500 n+204\big)}{3(n+1)^4 (n+2)^3} 
        +\frac{\zeta(2)}{(n+1) (n+2)}
\right)\\
&+\frac{2\big(
        18 n^5+150 n^4+490 n^3+755 n^2+536 n+132\big)}{3(n+1)^4 (n+2)^3}  
+\frac{(2 n+3) \zeta(2)}{(n+1)^2 (n+2)}\\
&+\left(
        -\frac{4}{(n+1)^2 (n+2)}
        +\frac{4 (-1)^n}{(n+1) (n+2)}
\right) S_2(n)\\
&+\left(
        \frac{4 (-1)^n}{3 (n+1) (n+2)}
        -\frac{4 (n+9)}{3 (n+1)^2 (n+2)}
\right) S_{-2}(n);
\end{align*}
here $S_a(n)=\sum_{k=1}^n\frac{\text{sign(a)}^k}{k^a}$ are the generalized harmonic numbers and $\zeta(z)=\sum_{k=1}^{\infty}\frac{1}{k^z}$ denotes the Riemann zeta function. 
\end{example}

\noindent As illustrated in the previous example, one can combine the Almkvist-Zeilberger method with the above recurrence solver to obtain a method\footnote{If the integrand has a particularly nice shape (e.g., if it is proper hyperexponential in the $x_i$ and proper hypergeometric in $n$), the multivariate Almkvist Zeilberger method will provide such a recurrence~\eqref{Equ:RecEpFull} with $b(n,\ep)=0$. In short, for such special input the method under consideration turns to a complete decision procedure.} for the following problem.

\medskip

\ProblemSpec{\noindent \textbf{Simplifying multiple integrals.}\\
\textit{Given} an integral in the form~\eqref{Equ:GeneralForm} 
where $f$ is hyperexponential in $x_i$ for $i=1,2,\dots,s$ and hypergeometric in $n$.
\textit{Compute} the first coefficients of its $\ep$-expansion in terms of indefinite nested sums over hypergeometric products.}

\subsection{Symbolic summation: the difference ring approach}\label{Sec:Summation}

Given a sum representation of the form~\eqref{Equ:SumRep}, symbolic summation algorithms in the setting 
of difference rings and fields~\cite{Karr:81,Abramov:94,Singer:99,Schneider:01,Schneider:05c,Schneider:07d,Schneider:08c,Petkov:10,Schneider:10b,Schneider:15,DR1,DR2,DR3} can be 
utilized to derive an alternative representation in terms of indefinite nested sums over hypergeometric 
products. 

\medskip

\ProblemSpec{\textbf{Simplifying multiple sums.}\\
\textit{Given} a multiple sum of the form~\eqref{Equ:SumRep} where $f$ is hypergeometric in $n$ and $k_i$ for $i=1,2,\dots,v$. \textit{Find} a simplified version in terms of indefinite nested sums over hypergeometric products.}

\medskip

\noindent The proposed method works from inside to outside of such a multiple sum and transforms in each step the arising definite summation quantifier to an indefinite nested version by using the 
Mathematica package \texttt{Sigma}~\cite{Schneider:07a,Schneider:13a}. Namely, suppose that we transformed already a sub-sum of~\eqref{Equ:SumRep} to an expression in terms of indefinite nested sums w.r.t.\ $k$ yielding the expression $f(\dots,m,k)$. Moreover, suppose that we have to deal with an extra summation quantifier, say with $F(\dots,m)=\sum_{k=0}^{L(\dots,m)}f(\dots,m,k)$. If this summation quantifier is the outermost sum in~\eqref{Equ:SumRep}, then $m$ is precisely the Mellin variable $n$. Otherwise, $m$ will be the summation variable of the next summation quantifier that is applied to $F(\dots,m)$. Then the following three steps lead often to the 
desired simplification. 
\begin{enumerate}
 \item Given the definite sum $F(m)$ (we suppress further variables), try to compute a linear recurrence relation of the form~\eqref{Equ:RecNoEp} ($n$ replaced by $m$) using the creative telescoping 
paradigm~\cite{AequalB} in the setting of difference rings~\cite{Schneider:08c,Schneider:15,DR1,DR2}. Here $a_i(m)$ are polynomials in $m$ and $b(m)$ is an expression in terms of indefinite nested sums over hypergeometric products.
\item Solve afterwards the recurrence in terms of indefinite nested sums over hypergeometric products w.r.t.\ $m$ using the toolbox described in Section~\ref{Sec:RecurrenceSolving}. 
\item Finally, try to combine\footnote{This will works in general if one finds $d$ linearly independent solutions of the homogeneous version of the recurrence and one particular solution of the recurrence itself.} the solutions to derive a simpler representation of the original input sum in terms of indefinite nested sums over hypergeometric products w.r.t.\ $m$.
\end{enumerate}

\noindent Suppose that all three steps can be carried out. If $n=m$ is the Mellin parameter in~\eqref{Equ:SumRep} then we are done. Otherwise, $F(\dots,m)$ will take over the role of $f(\dots,k)$ and we repeat this game to treat the next summation quantifier in~\eqref{Equ:SumRep}.\\
In some rare cases, this machinery even works if the input sum depends on $\ep$. However, in most cases one will arrive at linear recurrences that depend on $\ep$ and that does not have sufficiently many solutions in terms of indefinite nested sums. As a consequence, step 3 from above will fail. A very successful strategy is based on calculating an $\ep$-expansion of the summand of~\eqref{Equ:SumRep} up to the desired order and applying\footnote{If infinite sums are involved, properties such as uniformal convergence have to be taken extra into account.} the summation quantifiers to each of the coefficients (which are free of $\ep$). Afterwards, the summation mechanism described above is applied to each of the arising summation problems. In many applications, the derived recurrences (now free of $\ep$) are completely solvable in terms of indefinite nested sums and all three steps 1--3 from above can be carried out successfully.
The package \texttt{EvaluateMultiSums}~\cite{Schneider:13a,Schneider:13b} based on \texttt{Sigma} combines all these steps and variants thereof in a very efficient way in order to carry out such simplifications automatically.

\begin{example}
Consider the 2-mass 3-loop Feynman diagram\footnote{The graph has been drawn using {\tt Axodraw} \cite{Vermaseren:1994je}.}
\begin{equation}\label{Equ:AggDiagram}
\includegraphics[width=3cm]{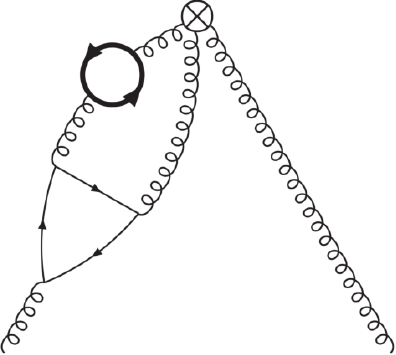},
\end{equation}
that arose in the calculation of the gluonic operator matrix element $A_{gg,Q}^{(3)}$~\cite{2Mass3Loop:18}. The corresponding integral is transformed to an expression of size 95MB given as a linear combination of multiple sums of the form~\eqref{Equ:SumRep}. More precisely, 
the expression is built by 150 single sums, 1000 double sums, 12160 triple sums and 1555 quadruple sums.
A typical triple sum is
\begin{multline*}
\sum_{j=0}^{n}\sum_{i=0}^j\sum_{k=0}^{i}\frac{(4+\ep) (-2+n) (-1+n) n \pi (-1)^{2-k}}{2+\ep}\times 2^{-2+\ep} e^{-\frac{3 \ep \gamma }{2}} \eta^{k}\times\\[-0.2cm]
\times\tfrac{        \Gamma (
                1
                -\frac{\ep}{2}
                -i
                +j
                +k
        )\Gamma (
        -1-\frac{\ep}{2}) \Gamma (
        2+\frac{\ep}{2}) \Gamma (1+n) \Gamma (1
+\ep
+i
-k
) \Gamma (
        -\frac{3 \ep}{2}
        +k
) \Gamma (1
-\ep
+k
) \Gamma (3
-\ep
+k
) \Gamma (
        -\frac{1}{2}
        -\frac{\ep}{2}
        +k
)}{\Gamma (
        -\frac{3}{2}-\frac{\ep}{2}) \Gamma (
        \frac{5}{2}+\frac{\ep}{2}) \Gamma (2+i) \Gamma (1+k) \Gamma (2
-i
+j
) \Gamma (2
-\ep
+k
) \Gamma (
        \frac{5}{2}
        -\ep
        +k
) \Gamma (
        -\frac{\ep}{2}
        +k
) \Gamma (
        5
        +\frac{\ep}{2}
        +n
)}
\end{multline*}
where $\eta=\frac{m_1}{m_2}$ is the quotient of the two arising masses.
Applying the above tools for the $\ep^{-1}$-contribution yields a simplification in terms of indefinite nested sums in about 30 minutes. This suggests that the full $\ep^{-1}$ contribution of the above diagram will require about 1 year of calculation time. Interestingly enough, our summation tools failed to produce the $\ep^0$ contribution. More precisely, we obtained recurrences that could not be solved in terms of indefinite nested sums. \\
As a consequence, we used the package \texttt{SumProduction} to crunch the derived expression consisting of 14865 sums to an expression consisting only of 8 sums (which are rather large). Afterwards, we applied our summation tools to these 8 sums. The calculations can be summarized in the following table~\cite{2Mass3Loop:18}.
\small
$$\begin{tabular}{|r|r|r|rr|r|}
\hline
sum & size of sum & summand size of & time of & & number of\\
&(with $\ep$)& constant term &calculation&&indef.\ sums\\
\hline
\hline
{\tiny$\displaystyle\sum_{i_4=2}^{n-3}\sum_{i_3=0}^{i_4-2}\sum_{i_2=0}^{i_3}\sum_{i_1=0}^{\infty}$} &  17.7 MB & 266.3 MB & 177529 s & (2.1 days) & 1188\\
{\tiny$\displaystyle\sum_{i_3=3}^{n-4}\sum_{i_2=0}^{i_3-1}\sum_{i_1=0}^{\infty}$} & 232 MB &  1646.4 MB &980756 s & (11.4 days) & 747\\
{\tiny$\displaystyle\sum_{i_2=3}^{n-4}\sum_{i_1=0}^{\infty}$}   & 67.7 MB & 458 MB & 524485 s & (6.1 days) & 557\\
{\tiny$\displaystyle\sum_{i_1=0}^{\infty}$} & 38.2 MB & 90.5 MB &689100 s & (8.0 days) & 44\\
\hline
{\tiny$\displaystyle\sum_{i_4=2}^{n-3}\sum_{i_3=0}^{i_4-2}\sum_{i_2=0}^{i_3}\sum_{i_1=0}^{i_2}$} & 1.3 MB & 6.5 MB & 305718 s & (3.5 days) & 1933\\
{\tiny$\displaystyle\sum_{i_3=3}^{n-4}\sum_{i_2=0}^{i_3-1}\sum_{i_1=0}^{i_2}$} & 11.6 MB & 32.4 MB & 710576 s & (8.2 days) & 621\\
{\tiny$\displaystyle\sum_{i_2=3}^{n-4}\sum_{i_1=0}^{i_2}$}   & 4.5 MB & 5.5 MB & 435640 s  & (5.0 days) & 536\\
{\tiny$\displaystyle\sum_{i_1=3}^{n-4}$} & 0.7 MB & 1.3 MB & 9017s & (2.5 hours) & 68\\
\hline
\end{tabular}$$
\normalsize
For instance, consider the quadruple sum of the form $\sum_{i_4=2}^{n-3}\sum_{i_3=0}^{i_4-2}\sum_{i_2=0}^{i_3}\sum_{i_1=0}^{\infty}$. In total the summand requires 17.7MB of memory. Taking its $\ep$-expansion, the constant term requires 266.3 MB. Applying afterwards the package \texttt{EvaluateMultiSums} to the quadruple sum of the constant term yields within 2.1 day an expression in terms of 1188 indefinite nested sums. In order to treat all 8 sums, we needed 3 month of calculation time. In total we obtain an expression for diagram~\eqref{Equ:AggDiagram} that uses 154MB of memory and is composed by 4110 indefinite nested sums.
For instance one of the most complicated sums is

\vspace*{-0.8cm}

\small
\begin{multline*}
\sum_{h=1}^n 2^{-2 h}(\text{$1-{\eta}$})^h \binom{2 h}{h} \left(
        \sum_{i=1}^h \tfrac{2^{2 i}(
                1-{\eta})^{-i}}{i \binom{2 i}{i}}
\right)\left(
        \sum_{i=1}^h \tfrac{(
                1-{\eta})^i \binom{2 i}{i}}{2^{2 i}}\right) 
\sum_{i=1}^h \frac{\displaystyle 2^{2 i} (
        1-{\eta})^{-i}\sum_{j=1}^i \frac{\displaystyle\sum_{k=1}^j\tfrac{(
        1-{\eta})^k}{k}}{j}}{i\binom{2 i}{i}}.
\end{multline*}
\normalsize
Finally, one can simplify this expression further by using the available difference ring algorithms of \texttt{Sigma}. The final result is an expression of only 8.3 MB size in terms of 74 binomial sums that are all algebraically independent among each other. 
\end{example}

\begin{remark}
We also started to explore the usage of holonomic summation tools~\cite{Chyzak:00,Schneider:05d} in the setting of QCD. We strongly expect that new ideas of~\cite{DRHolonomic} will push forward the available summation techniques.
\end{remark}

\subsection{Solving coupled systems and the large moment machinery}\label{Sec:CoupledSystem}
After applying IBP methods~\cite{Chetyrkin:1981qh,Laporta:2001dd,Studerus:2009ye,vonManteuffel:2012np,MARSEID} 
physical quantities are crunched to expressions in terms of master integrals. In particular, these master integrals are described as  solutions of coupled systems of linear difference or differential equations. E.g., in the univariate differential equation case they are of the form  
\begin{eqnarray}
\label{eq:DEQ1}
D_x
\left(
\begin{matrix}
f_1(x,\ep)\\ \vdots \\ f_{\lambda}(x,\ep)\end{matrix}\right) 
= A
\left(\begin{matrix} f_1(x,\ep)\\ \vdots \\ f_{\lambda}(x,\ep)\end{matrix}\right) 
+ \left(\begin{matrix} g_1(x,\ep)\\ \vdots \\ g_{\lambda}(x,\ep)\end{matrix}\right),
\end{eqnarray}
where $A$ is a $\lambda\times\lambda$ matrix with entries from $\KK(x,\ep)$ and the entries of the right-hand side vector $g(x,\ep)=(g_1(x,\ep),\dots g_{\lambda}(x,\ep))$ are given as a linear combination of simpler master integrals $h_1(x,\ep),\dots,h_u(x,\ep)$ over $\KK(x,\ep)$. They can be either determined by other coupled systems (by using the proposed method recursively) or have to be tackled, e.g., by our symbolic summation and integration tools introduced above. In the following we suppose that the unknown functions $f_i(x,\ep)$ have a power series representation
\begin{equation}\label{Equ:FiFormalP}
f_i(x,\ep)=\sum_{k=0}^{\infty}F_{i}(k,\ep)x^k,
\end{equation}
and we seek for symbolic representations of the coefficients of the $\ep$-expansions
\begin{equation}\label{Equ:FiLaurenS}
F_{i}(k,\ep)=\sum_{j=l}^{\infty}F_{i,j}(k)\ep^j
\end{equation}
with $1\leq i\leq\lambda$.
Likewise we assume that the simpler master integrals $h_i(x,\ep)$ with $1\leq i\leq u$
arising in $g(x,\ep)$ have power series representations whose coefficients can be represented in terms of indefinite nested sums. More precisely, their $\ep$-expansions are assembled by coefficients that can be expressed in terms of indefinite nested sums. Then one can utilize the following algorithm that is implemented within the package \texttt{SolveCoupledSystem}~\cite{CoupledSys:15,Schneider:16b} and is based on \texttt{Sigma}.

\medskip

\ProblemSpec{\textbf{Solving coupled systems.}\\ 
\textit{Given} a system as above~\eqref{eq:DEQ1} with initial values of the desired solution and $l,r\in\ZZ$ with $l\leq r$.
\textit{Decide} constructively if the coefficients $F_{i,j}(k)$ for $i=1,\dots,\lambda$, $j=l,\dots,r$ 
of the power series $F_i(k,\ep)$ in~\eqref{Equ:FiFormalP} which form the coefficients of the $\ep$-expansions~\eqref{Equ:FiLaurenS} can be expressed in terms of indefinite nested sums.}
\begin{enumerate}
 \item Uncouple the system by using, e.g., Z\"urcher's algorithm~\cite{Zuercher:94,BCP13} implemented in the package 
\texttt{OreSys}~\cite{ORESYS}. Usually\footnote{In general one might obtain several scalar linear differential equations. Then the described steps in 2--4 below are carried out for each equation.} one gets one scalar linear differential equation in one of the unknown functions, say $f_1(x,\ep)$ where the right-hand side can be given in terms of a power series whose coefficients are given explicitly in terms of indefinite nested sums. In addition, the remaining functions $f_2(x,\ep),\dots,f_{\lambda}(x,\ep)$ can be expressed by $f_1(x,\ep)$ and the simpler master integrals $h_i(x,\ep)$ with $1\leq i\leq u$ in the following form:
 \begin{equation}\label{Equ:fiRep}
 f_i(x,\ep)=\sum_r \alpha_{i,r}(x,\ep)D_x^rf_1(x,\ep)+\sum_{j=1}^u\sum_r \beta_{i,j,r}(x,\ep)D_x^rh_j(x,\ep)
 \end{equation}
 for some rational functions $\alpha_{i,r}(x,\ep),\beta_{i,j,r}(x,\ep)\in\KK(x,\ep)$.
 \item By holonomic closure properties~\cite{KP:11} compute a linear recurrence of $F_1(k,\ep)$ from the given scalar differential equation of $f_1(x,\ep)$. The recurrence is of the form~\eqref{Equ:RecEpFull} where the coefficients in~\eqref{Equ:RHSExpansion} can be given explicitly in terms of indefinite nested sums.
 \item Use the tools from Section~\ref{Sec:RecurrenceSolving} with the corresponding initial values in order to decide algorithmically if the first coefficients $F_{1,j}(k)$ for $j=l,\dots,r$ can be given in terms of indefinite nested sums. 
  If this is not possible, return that such a representation is not possible.
 \item Otherwise, plug these coefficients into~\eqref{Equ:FiLaurenS} (for $i=1$) and afterwards plug it into~\eqref{Equ:fiRep}. Similarly, proceed to plug the known coefficients of the simpler master integrals $h_i(x,\ep)$ with $1\leq i\leq u$ into~\eqref{Equ:fiRep}. Finally, extract the first coefficients of the $\ep$-expansions of $F_2(k,\ep),\dots,F_{\lambda}(k,\ep)$.
\end{enumerate}

\begin{remark}\label{Remark:OrderOfEp}
Often one has to calculate the $\ep$-expansions for $F_1(k,\ep)$ and for the simpler master integrals $h_j(x,\ep)$ up to a certain order that is higher than $r$ due to factors $\frac1\ep$. The corresponding bounds can be determined after the uncoupling has been carried our. We neglect further details on these technical aspects and refer to~\cite{CoupledSys:15}. 
\end{remark}

\noindent Using this toolbox (and slight variants of it~\cite{Bluemlein:2014qka,Schneider:16b,CoupledSys:15}) we performed already rather advanced QCD-calculations~\cite{CoupledSys1,CoupledSys2,CoupledSys3,Schneider:16b,CoupledSys4,CoupledSys5,Physics1,Physics2}.

\begin{remark}
In certain instances one can also use ideas from~\cite{Henn:2013pwa,Lee:2014ioa} to find solutions in terms of indefinite nested integrals. 
\end{remark}


An interesting feature is that this method can be adapted to calculate a large number moments~\cite{Blumlein:2017dxp}, say $\mu=8000$, that is also implemented within the package \texttt{SolveCoupledSystem}.

\medskip

\ProblemSpec{\textbf{Large moment method.}\\ 
\textit{Given} a system as above~\eqref{eq:DEQ1} with initial values of the desired solution, $l,r\in\ZZ$ with $l\leq r$ and $\mu\in\NN$.
\textit{Calculate efficiently} $F_{i,j}(0),F_{i,j}(1),\dots,F_{1,j}(\mu)$ for $i=1,\dots,\lambda$, $j=l,\dots,r$ of the power series $F_i(k,\ep)$ in~\eqref{Equ:FiFormalP} which form the coefficients of the $\ep$-expansions~\eqref{Equ:FiLaurenS}.\looseness=-1}
\begin{enumerate}
 \item 
Suppose that one has calculated already the first $\mu$ moments of the simpler master integrals that arise in the right-hand side of $g_i(x,\ep)$ up to a certain order. This can be accomplished, e.g., by applying this method again to this simpler integrals or by utilizing the representations in terms of indefinite nested sums coming from our symbolic summation and integrations tools described above.  As a consequence one can calculate the sequence of values $b_i(0),b_i(1),\dots,b_i(\mu)$ of~\eqref{Equ:RHSExpansion} in~\eqref{Equ:RecEpFull} up to a certain order, say $l\leq i\leq r$. 
\item  Using the recurrence~\eqref{Equ:RecEpFull} with the moments given on the right-hand side one can calculate in linear time the moments  
$F_{1,j}(0),F_{1,j}(1),\dots,F_{1,j}(\mu)$ for $j=l,\dots,r$ of the $\ep$-expansion~\eqref{Equ:FiLaurenS} ($i=1$) of $F_1(k,\ep)$ in~\eqref{Equ:FiFormalP} provided that $F_{1,j}(0),F_{1,j}(1),\dots,F_{1,j}(d)$ is given (using our summation tools or using from above or procedures like {\tt Mincer} \cite{Mincer:91} or {\tt MATAD} \cite{Steinhauser:2000ry}).
\item Finally, one plugs these values into~\eqref{Equ:fiRep} and extracts the values $F_{r,j}(0),F_{r,j}(1),\dots,F_{r,j}(\mu)$ for $r=2,\dots,\lambda$ and $j=l,\dots,r$. 
\end{enumerate}

\begin{remark}
The comments in Remark~\ref{Remark:OrderOfEp} are also here relevant to calculate the correct moments up to the desired $\ep$-order $r$. In addition, it might happen that the number of moments of $f_1(x,\ep)$ in calculation step~(2) and the moments of the simpler master integrals $h_j(x,\ep)$ must be chosen slightly higher than $\mu$ in order to get all moments correctly. This is owing to the fact that extra factors $x$ might occur (also during the uncoupling process) which introduce negative shifts on the coefficient level.
\end{remark}

\noindent Recall that the IBP methods usually crunch physical expressions to simper expressions in terms of the master integrals under consideration. Assembling all these moments produces large moments of the physical quantities. With these moments one is now in the position to guess, e.g., homogeneous recurrence relations 
using packages like~\cite{GSAGE}. Afterwards, one can use our recurrence solvers from Section~\ref{Sec:RecurrenceSolving} to derive representations in terms of indefinite nested sums.
First non-trivial  case studies have been carried out, like the first re-computation of the  3-loop splitting functions~\cite{CoupledSys5} and simpler parts of the massive 3-loop form factor~\cite{FORMF3}.

\section{Conclusion}

We presented the central methods for symbolic summation/integration, solving linear recurrences and solving coupled systems of linear differential equations that have been encoded within the Mathematica packages \texttt{Sigma}, \texttt{EvaluateMultiSum}, \texttt{MultiIntegrate}, \texttt{SumProduc\-tion}  and \texttt{SolveCoupledSystem}. Besides these computer algebra tools also special function algorithms~\cite{Vermaseren:99,Remiddi:1999ew,Bluemlein2009a,ABS:11,ABS:13,ABRS:14,AS:18} implemented within the package \texttt{HarmonicSums}~\cite{Ablinger:12} are used in order to speed up the above methods; in particular the limit $n\to\infty$ can be performed for expressions in terms of indefinite nested sums in $n$ by utilizing asymptotic expansion algorithms.\\ 
It will be interesting to see how all these computer algebra and special function algorithms can be generalized to tackle also more complicated special functions like iterative-non-iterative integrals and sums~\cite{Noniterative:17} as they appeared in recent calculations. In this regard, we expect that the large moment machinery (see Section~\ref{Sec:CoupledSystem}) will play a decisive role within our future calculations.


\providecommand{\noopsort}[1]{}

\end{document}